\documentclass[11pt,twoside]{article}
\usepackage[paperwidth={7.87in},paperheight={10.24in},left={0.87in},right={0.87in},top={1.3in},bottom={1.3in}]{geometry}
\usepackage{abstract}
\usepackage{fancyhdr}
\usepackage{url}
\usepackage{times}
\usepackage{graphicx}
\usepackage{amsmath}
\usepackage{amsfonts}
\usepackage{amssymb}
\usepackage{mathrsfs}
\usepackage[super,sort&compress]{natbib}
\usepackage{color}
\usepackage{lscape}
\usepackage{ulem}
\usepackage{titlesec}
\usepackage{tabularx}
\usepackage{rotating}
\usepackage{marginnote}
\usepackage{setspace}
\usepackage[labelfont=bf]{caption}

\usepackage{lineno}

\titleformat*{\section}{\large\bfseries}
\titleformat*{\subsection}{\large\itshape}

\newcommand*\thetitle{Channelization cascade}

\newcommand*\theauthors{Sara Bonetti$^{1}$, Milad Hooshyar$^{2,3}$, Carlo Camporeale$^{4}$, and Amilcare Porporato$^{2,3,\ast}$ \\
$^{1}${\small Department of Environmental Systems Science, ETH Zurich, 8092 Zurich, Switzerland} 
\\
$^{2}${\small Department of Civil and Environmental Engineering, Princeton University, Princeton, New Jersey 08544, USA} 
\\
$^{3}${\small Princeton Environmental Institute, Princeton university, Princeton, New Jersey 08544, USA}
\\
$^{4}${\small Department of Environment, Land and Infrastructure Engineering, Politecnico di Torino, 10129 Turin, Italy} 
\\
$^{\ast}${\small Corresponding author: aporpora@princeton.edu}
}
\newcommand*\runningauthor{Bonetti et al.}

\fancypagestyle{plain}{%
\fancyhf{}

\fancyhead[OR]{}
\fancyhead[OL]{}
\fancyhead[OC]{}
\fancyhead[ER]{}
\fancyhead[EL]{}
\fancyhead[EC]{}
}
\begin{document}
%
\raggedbottom
\title{\textsf {\Large \bfseries \thetitle}}
\author{\theauthors}
\date{}
\maketitle
\textbf{The hierarchy of channel networks in landscapes displays features that are characteristic of non-equilibrium complex systems. Here we show that a sequence of increasingly complex ridge and valley networks is produced by a system of partial differential equations coupling landscape evolution dynamics with a specific catchment area equation. By means of a linear stability analysis we identify the critical conditions triggering channel formation and the emergence of characteristic valley spacing.
The ensuing channelization cascade, described by a dimensionless number accounting for diffusive soil creep, runoff erosion, and tectonic uplift, is reminiscent of the subsequent instabilities in fluid turbulence, while the structure of the simulated patterns is indicative of a tendency to evolve toward optimal configurations, with anomalies similar to dislocation defects observed in pattern-forming systems.
The choice of specific geomorphic transport laws and boundary conditions strongly influences the channelization cascade, underlying the nonlocal and nonlinear character of its dynamics.
}


\vspace{5mm}

The spatial distribution of ridges and valleys is one of the most striking features of a landscape.
It has long fascinated the physical, geomorphological, and hydrological communities, leading to the development of a rich body of work on the statistical, theoretical, and numerical analysis of landscape organization. Early works focused on the definition of stream ordering systems for the river basin characterization \cite{Horton1945,Strahler1952,Shreve1966} and the coupled dynamics of water and sediment transport to identify stability conditions for incipient valley formation\citep{Smith1972,Loewenherz1991,Izumi1995}, followed by the statistical description of river networks, including scaling laws and fractal properties of river basins \cite{Tarboton1988,Marani1991,Rodriguez2001,Dodds2000}, the related optimality principles\cite{Rigon1993,Rodriguez2001}, and stochastic models \cite{Banavar1997,Somfai1997,Pastor1998}. These studies have shed light on the spatial organization and governing statistical laws of developed river networks and explored the linkages to other branch-forming systems \citep{Kramer1992,Arneodo1992,Somfai1997}, but have not tackled the physical origin of the underlying instabilities and feedback mechanisms acting over time in the formation of the observed ridge and valley patterns \citep{Fowler2011}. To this purpose, landscape evolution models have been employed for the analysis of branching river networks \citep{Perron2008,Perron2012} in relation to the main erosional mechanisms acting on the topography. These works represented an important step forward in the study of spatially organized patterns of ridges and valleys. However, lacking a rigorous formulation of the drainage area equation \citep{Gallant2011,Bonetti2018PRSA}, they resort to grid-size dependent algorithms \citep{Schoorl2000,Pelletier2012} coupled to empirical adjustments \cite{Perron2008,Pelletier2012}, thus precluding the theoretical investigation of the underlying instabilities.

In this work, we focus on landscapes characterized by runoff erosion, expressed as a function of the specific drainage area $a$ (ref. \citenum{Bonetti2018PRSA}) to obtain grid-independent solutions without the introduction of additional system parameters. The resulting system of coupled, nonlinear partial differential equations (PDEs) provides a starting point for the theoretical analysis of channel-forming instabilities and landscape self-organization. The nonlocal character of the equations makes the boundary conditions extremely important. On regular domains, simulations reveal a sequence of channel instabilities reminiscent of the laminar-to-turbulent transition \citep{Pope2000,Drazin2004,Kundu2011}. 
The explicit mathematical structure makes it possible to  perform a linear stability analysis of the coupled PDE system to identify the critical conditions for the first channel-forming instability.
The subsequent branching sequence towards smaller and smaller valleys until soil creep becomes dominant is similar to the turbulent cascade with large scale vortices leading to smaller ones until viscous dissipation. The formation of regular pre-fractal networks of ridges and valleys, brought about by the regular boundary conditions, also reveals the tendency of the system to develop towards optimal configurations suggestive of maximization principles\cite{Rigon1993} typical of non-equilibrium thermodynamics \citep{Rinaldo1996,Ozawa2003,Martyushev2006}, complex branching systems \citep{Arneodo1992,Bensimon1986,Sander2005}, and in particular of constructal theory \citep{Errera1998,Lorente2002,Bejan2011,Bejan2016}. Our analysis is different from recent interesting contributions on groundwater-dominated landscapes \citep{Devauchelle2012,Yi2017}, where branching and valley evolution is initiated at seepage points in the landscape.

\section*{Landscape evolution in detachment-limited conditions}
The time evolution of the surface elevation $z(x,y,t)$ is described by the sediment continuity equation \citep{Dietrich2003,Perron2008,Smith2010,Fowler2011}
\begin{equation}
	\frac{\partial z}{\partial t}=U-\nabla \cdot \mathbf{f}=U-\nabla \cdot \left(\mathbf{f_d}+\mathbf{f_c}\right),
	\label{eq5:z1}
\end{equation}
where $t$ is time, $U$ is the uplift rate, and $\mathbf{f}$ is the total sediment flux, given by the sum of fluxes related to runoff erosion/channelized flow ($\mathbf{f_c}$) and soil creep processes ($\mathbf{f_d}$). The soil creep flux is assumed to be proportional to the topographic gradient \citep{Culling1960,Culling1963}, hence $\mathbf{f_d}=-D \nabla z$, $D$ being a diffusion coefficient (here assumed to be constant in space and time). In the so-called detachment-limited (DL) conditions \citep{Howard1994,Izumi1995,Perron2008} it is assumed that all eroded material is transported outside the model domain, so that no sediment redeposition occurs. Under these conditions, the runoff erosion term is approximated as a sink term given by \citep{Perron2008} $\nabla \cdot \mathbf{f_c}= K'_a |\nabla z|^n q^m$ , where $K'_a$ is a coefficient, $q$ is the discharge per unit width of contour line, and $m$ and $n$ are model parameters. As a result, equation \eqref{eq5:z1} becomes
\begin{eqnarray}
	\frac{\partial z}{\partial t}= D \nabla^2 z - K'_a q^m |\nabla z|^n +U.
	\label{eq5:z2}
\end{eqnarray}
Thus the soil creep flux results in a diffusion term which tends to smooth the surface, while the runoff erosion component is a sink term which excavates the topography as a function of local slope and specific water flux.

The surface water flux $q$ is linked to the continuity equation
\begin{eqnarray}
	\frac{\partial h}{\partial t}= R - \nabla \cdot (q \mathbf{n})
	\label{eq5:h}
\end{eqnarray}
where $h$ is the water height, $\mathbf{n}$ the direction of the flow, and $R$ the rainfall rate effectively contributing to runoff production. Equation \eqref{eq5:h} can be simplified assuming steady-state conditions with constant, representative rainfall rate, $R_0$, and (as in previous works\citep{Rodriguez1992b}) constant speed of water flow $v_0$ in the direction opposite to the landscape gradient (i.e., $\mathbf{n}=-\nabla z/|\nabla z|$). In such conditions, it can be shown \citep{Bonetti2018PRSA} that the water height, $h$, and the specific water flux, $q$, are both proportional to the specific contributing area, $a$, i.e. $h=q/v_0=a R_0/v_0$. As a result, the system of Equations \eqref{eq5:h} - \eqref{eq5:z2} reduces to an equation for the specific catchment area $a$ (see ref. \citenum{Bonetti2018PRSA}),
\begin{eqnarray}
	-\nabla \cdot \left(a \frac{\nabla z}{|\nabla z|}\right)=1,
	\label{eq5:a}
\end{eqnarray}
coupled to the landscape evolution equation
\begin{eqnarray}
	\frac{\partial z}{\partial t}= D \nabla^2 z - K_a a^m |\nabla z|^n +U,
	\label{eq5:z3}
\end{eqnarray}
 with an adjusted erosion constant $K_a$ to account for the proportionality between $a$ and $q$.

It is important to observe that the specific drainage area $a$ has units of length and is related to the drainage area $A$ as $a=\lim_{w \to 0}A/w$; it is thus defined per unit width of contour line $w$ (see ref. \citenum{Bonetti2018PRSA}). Most landscape evolution models (see, e.g., refs. \citenum{Rodriguez2001,Perron2008,Pelletier2012,Chen2014}) use the total drainage area $A$ in equation \eqref{eq5:z3} instead of $a$, with several notable implications. The value of $A$ is generally evaluated using numerical flow-routing algorithms (e.g., D8, D$\infty$ -- ref. \citenum{Tarboton1997}) which provide grid-dependent values of $A$.
To correct for this, the drainage area $A$ is then often modified to account for the channel width\cite{Perron2008,Pelletier2012}, but this results in approximations with arbitrary parameters. Conversely, the use of $a$ avoids grid-dependence of the resulting topography.  Moreover, re-casting the problem in terms of a consistent coupled system of PDEs makes it possible to analyze theoretically the landscape evolution process. As detailed below (see Methods), an analytic solution for the steady state hillslope profile can be derived and then used as a basic state for a linear stability analysis to identify the critical conditions for the first channel formation and the characteristic valley spacing.

It is useful to non-dimensionalize the system of equations \eqref{eq5:a} and \eqref{eq5:z3} to quantify the relative impact of soil creep, runoff erosion, and uplift on the landscape morphology. Using a typical length scale of the domain, $l$, and the parameters of equations \eqref{eq5:a} and \eqref{eq5:z3}, the following dimensionless quantities can be introduced: $\hat{t}= \frac{t D}{ l^2}$, $\hat{x}=\frac{x}{l}$, $\hat{y}=\frac{y}{l}$, $\hat{z}= \frac{z D}{U l^2 }$, and $\hat{a}= \frac{a}{l}$. With these quantities, equation \eqref{eq5:z3} becomes
\begin{equation}
	\frac{\partial \hat{z}}{\partial \hat{t}}= \hat{\nabla}^2 \hat{z}-\chi \hat{a}^m | \hat{\nabla}\hat{z}|^n+1
\end{equation}
where
\begin{equation}
	\chi= \frac{K_a l^{m+n}}{D^n U^{1-n}}.
	\label{eq5:chi}
\end{equation}
As we will see later, this describes the tendency to form channels in a way which is reminiscent of the global Reynolds number in fluid mechanics, as well as of the ratio of flow permeabilities used in constructal theory \cite{bejan2008design}. A similar quantity based on a local length scale (i.e., the mean elevation of the emerging topographic profile) was used in Perron et al.\cite{Perron2008}. The definition of $\chi$ as a function of global variables based on system parameters (e.g., uplift rate $U$) and boundary conditions allows us to directly infer system behavior. For example, when the slope exponent $n$ is equal to 1, the relative proportion of runoff erosion and soil creep can be seen to be independent of the uplift rate; however, if $n>1$ the uplift acts to increase the runoff erosion component, while for $n<1$ it enhances the diffusive component of the system. As we will see, this results in different drainage-network patterns as a function of uplift rates.

\begin{figure}
\centering
	\includegraphics[trim={0.5cm 0.5cm 0.5cm 1cm},width=\textwidth]{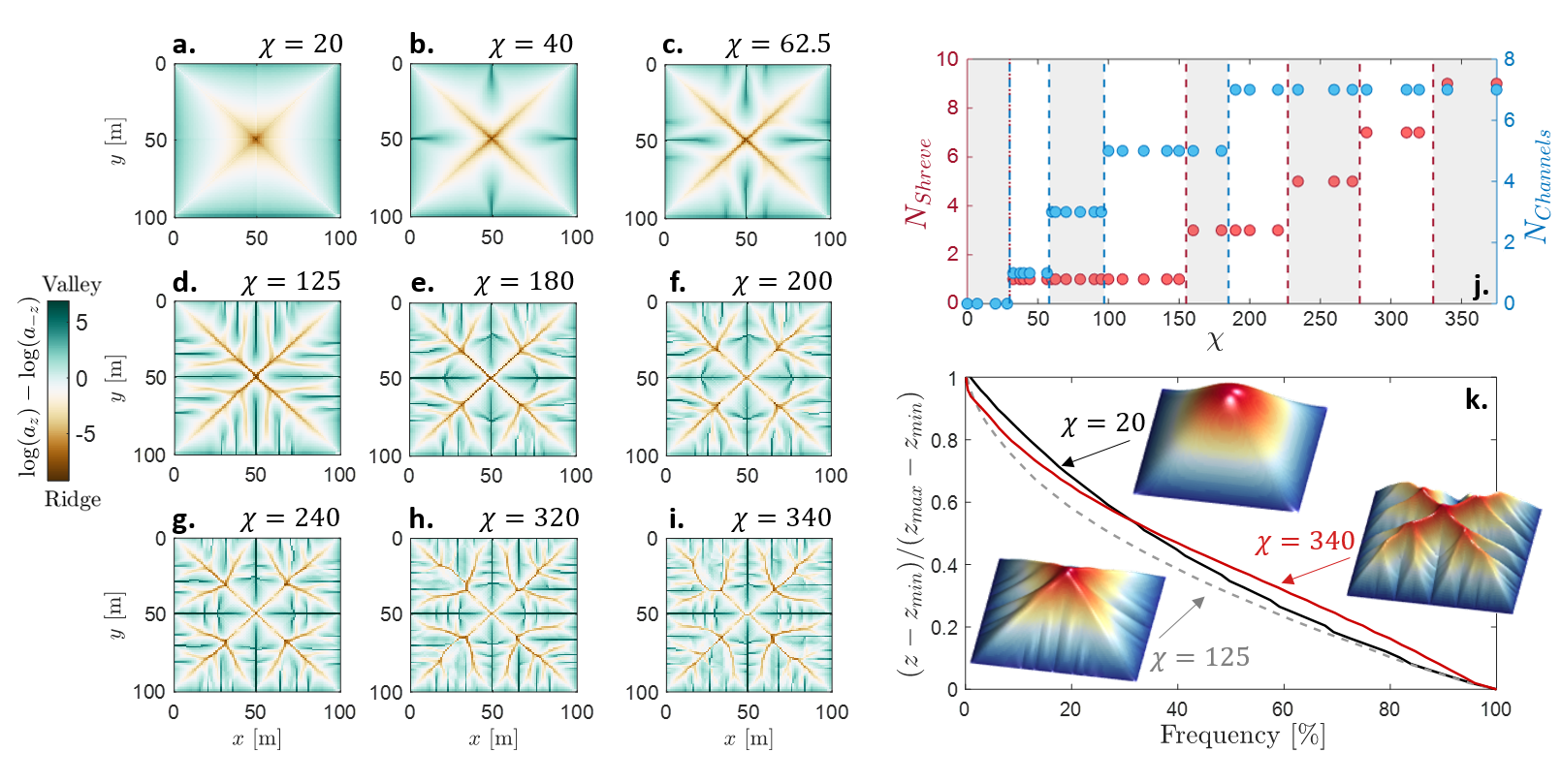}
	\caption{\textbf{Channelization cascade.} Simulation results obtained for $m=0.5$ and $n=1$. (a-i) Ridge and valley patterns obtained for $\chi$ values equal to 20, 40, 62.5, 125, 180, 200, 240, 320, and 340: brown corresponds to ridge lines and green to valleys (to better highlight the ridge and valley structure we show here the difference between the specific drainage area $a$ and its value computed over the flipped topography $-z$). (j) Highest Shreve order (red) and number of main channels on each domain side (blue) for different values of the dimensionless parameter $\chi$. Based on the number of channels and the Shreve order nine regimes can be identified with distinctively different ridge/valley patterns (shown  in panels a-i). (k) Normalized hypsometric curves obtained for $\chi$ =  20 (solid black), 125 (dashed gray), and 340 (solid red): when no secondary branching is observed  (i.e., $\chi \lesssim 155$) the hypsometric curve is concave, while after the first secondary branching is formed it undergoes a transition to a shape concave for higher elevations and convex at low elevations. Insets in panel k show 3d plots of the steady state topographies for the three cases, the color code represents surface elevation (red = high, blue = low).}
	\label{fig:results}
\end{figure}
\section*{Results}
\paragraph{Organized ridge and valley patterns.}
Simulation results obtained by numerically solving equations \eqref{eq5:a}-\eqref{eq5:z3} over square domains with $m=0.5$ and $n=1$ (see Methods for details) are shown in Fig. \ref{fig:results}. The emerging ridge/valley patterns are classified in terms of Shreve order (used here as a measure of branching complexity), and number of channels formed on each side of the domain. As can be seen from equation \eqref{eq5:chi}, for $n=1$ the dimensionless parameter $\chi$ is independent of the uplift rate, so that the spatial patterns of Fig. \ref{fig:results} are only a function of the relative proportions of the soil creep and runoff erosion components. For low $\chi$ values (i.e., $\lesssim 30$) no channels are formed and the topography evolves to a smooth surface dominated by diffusive soil creep (Fig. \ref{fig:results}a). As the runoff erosion coefficient is increased the system progressively develops one, three, and five channels on each side of the square domain for $30 \lesssim \chi \lesssim 58$, $58 \lesssim \chi \lesssim 97$, and $97 \lesssim \chi \lesssim 155$, respectively (Fig. \ref{fig:results}b-d). When $\chi$ is increased above $\approx 155$ the central channels develop secondary branches, with the main central channel becoming of Shreve order three (Fig. \ref{fig:results}e). As $\chi$ is further increased seven channels can be observed originating on each side of the domain, and the main central channel further branches (Fig. \ref{fig:results}f-i) becoming of order nine for the highest $\chi$ used for this configuration.

As the resulting landscape changes from a smooth topography to a progressively more dissected one, the shape of the hypsometric curve varies from concave (i.e., slope decreases along the horizontal axis) to one with a convex portion for low elevations (Fig. \ref{fig:results}k). In particular, channel formation (with no secondary branching) causes the hypsometric curve to progressively lower as a result of the lower altitudes observed in the topography, while maintaining a concave profile. As secondary branches develop, the hypsometric curve shifts to a concave/convex one, with the convex portion at lower altitudes becoming more evident as $\chi$ is increased (see red line for $\chi=340$ in Fig. \ref{fig:results}k).

The striking regularity of the drainage and ridge patterns induced by the simple geometry of the domain is reminiscent of regular pre-fractal structures (e.g., Peano basin \citep{Mandelbrot1982,Marani1991,Rodriguez1992,Flammini1996,Rodriguez2001}) and is indicative of the fundamental role of boundary conditions due to the highly non-local control introduced by the drainage area term. Particularly, the ridge and valley networks of Fig. \ref{fig:results} highly resemble Fig. 5 in ref. \citenum{Lorente2002}, where optimized tree-shaped flow paths were constructed to connect one point to many points uniformly distributed over an area. We further highlight similarities with the patterns obtained in ref. \citenum{Errera1998} by means of an erosion  model where the global flow resistance is minimized.

\begin{figure*}
	\centering
	\includegraphics[trim={0cm 0.5cm 0cm 0.5cm},width=\textwidth]{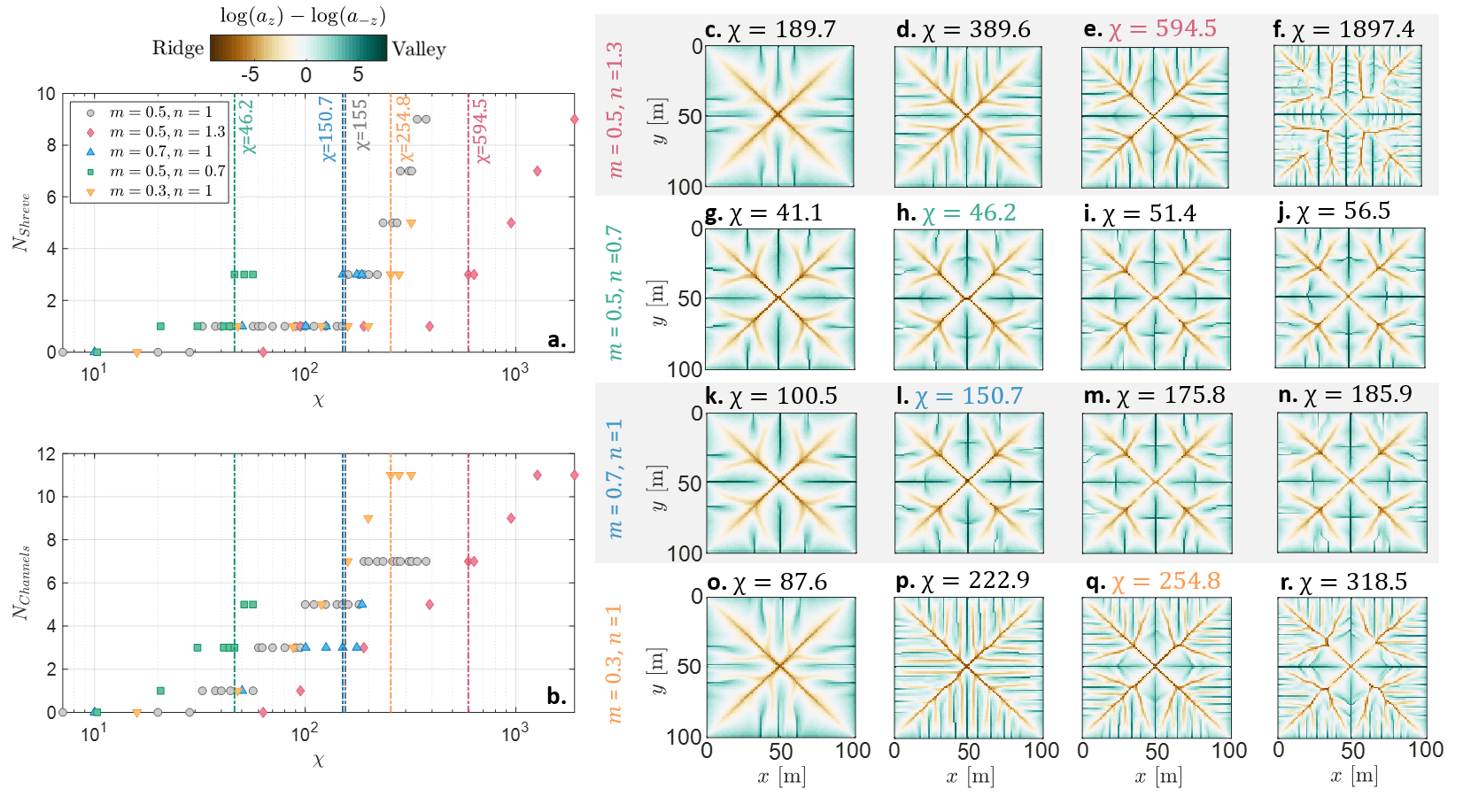}
	\caption{\textbf{Effect of runoff erosion laws.} Simulation results obtained for different values of the slope and runoff exponents (i.e., $n$ and $m$): (a) maximum Shreve order and (b) number of channels on each domain side as a function of $\chi$. Colored dash-dotted lines mark the $\chi$ values at which the first secondary branching is observed for each set of $m$ and $n$ values, and the corresponding ridge/valley patterns are highlighted in panels c-r. (c-r) Examples of two-dimensional ridge (brown) and valley (green) patterns for scenarios with (c-f) increased slope exponent ($n=1.3$, $m=0.5$, and $\chi=$ 189.7, 389.6, 594.5, 1897.4), (g-j) decreased slope exponent ($n=0.7$, $m=0.5$, and $\chi=$ 41.1, 46.2, 51.4, 56.6), (k-n) increased water flux exponent ($n=1$, $m=0.7$, and $\chi=$ 100.5, 150.7, 175.8, 185.9), and (o-r) decreased water flux exponent ($n=1$, $m=0.3$, and $\chi=$ 87.6, 222.9, 254.8, 318.5).}
	\label{fig:change_expon}
\end{figure*}
\paragraph{Effect of runoff erosion laws.}
The effect of different runoff erosion laws has been discussed in the literature \citep{Chen2014} also in relation to climate, vegetation cover, and soil properties \citep{montgomery2001climate,lowman2014investigating}.
Their role was analyzed here by changing the values of the exponents $n$ and $m$, as shown in Fig. \ref{fig:change_expon}.  

When the value of $n$ is different from unity, the resulting ridge/valley patterns depend on the uplift rate $U$, as per equation \eqref{eq5:chi}. When $n$ is increased the system displays channelization and secondary branching for higher values of $\chi$ (i.e., points are shifted to the right in Fig. \ref{fig:change_expon}a,b), with a more dissected planar geometry characterized by narrower valleys and smaller junction angles (Fig. \ref{fig:change_expon}c-f). A decrease in  $n$ leads to smoother geometries with wider valleys and the first secondary branching developing when only three channels per each side of the domain are present (see Fig. \ref{fig:change_expon}g-j). This results in a hypsometric curve with a more pronounced basal convexity as $n$ is increased above unity, as a consequence of the progressively more dissected topography (see SI, Fig. \ref{figS3:changeexponentshypso}).

When $m$ is increased (Fig. \ref{fig:change_expon}k-n) the system develops secondary branching when only three channels are present on each side of the domain, with the formation of less numerous but wider valleys with higher junction angles, and a reduced basal convexity in the hypsometric curve (Fig. \ref{figS3:changeexponentshypso}). Conversely, a decrease in $m$ results in a more dissected landscape, with narrower valleys (Fig. \ref{fig:change_expon}o-r) and a more pronounced transition of the hypsometric curve to a convex shape for low altitudes (Fig. \ref{figS3:changeexponentshypso}).

\begin{figure}
	\centering
	\includegraphics[trim={0.5cm 0.5cm 0.5cm 0cm},width=\textwidth]{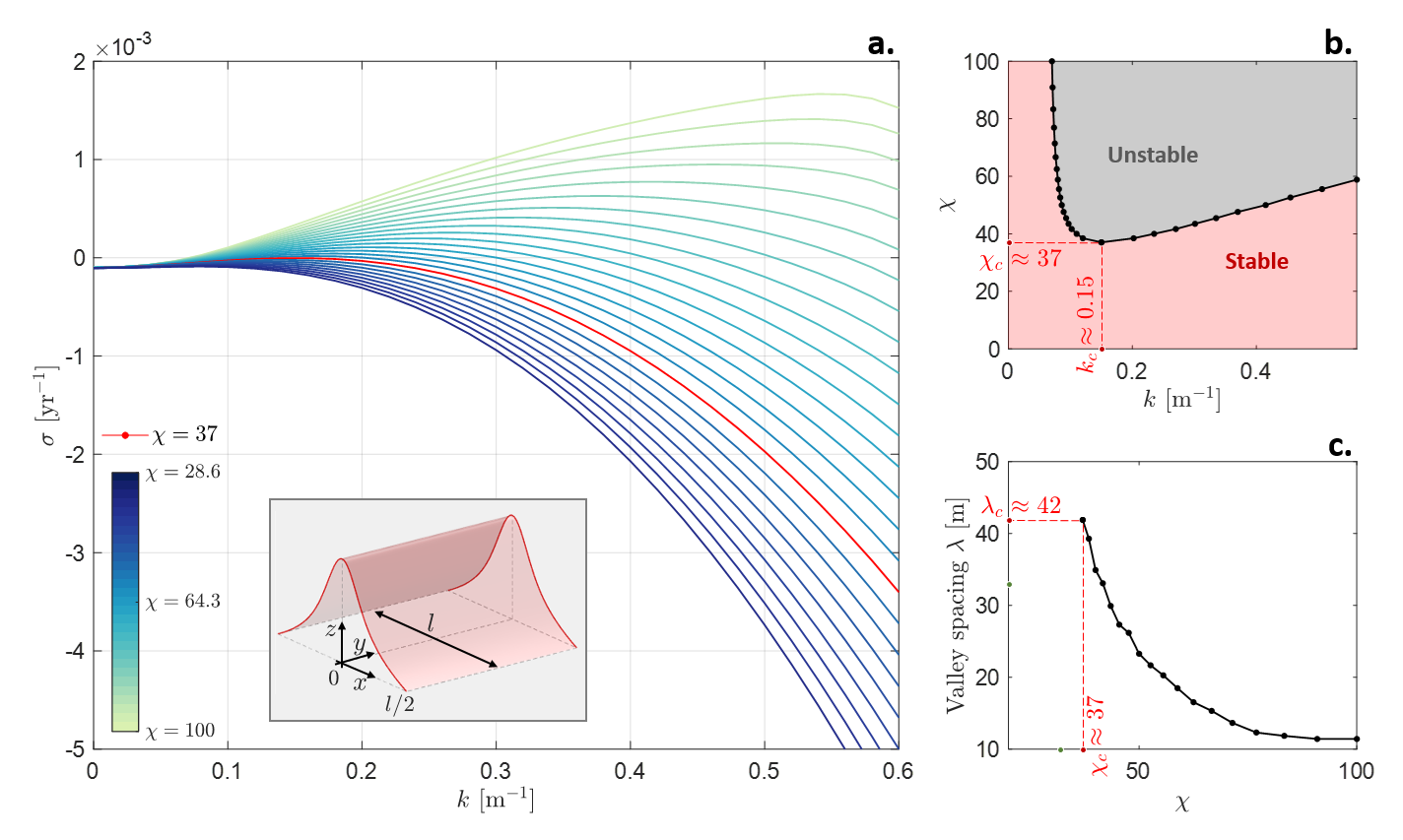}
	\caption{\textbf{Linear stability analysis.} (a) Growth rate $\sigma$ as a function of wavenumber $k$ for different values of the dimensionless number $\chi$, (b) marginal stability curve (the solid line marks the instability of the basic state to channel initiation), and (c) characteristic valley spacing $\lambda$ as a function of the dimensionless number $\chi$. The linear stability analysis predicts a critical value $\chi_c \approx 37$ for the first channel instability (with valley spacing $\lambda_c \approx 42$). 
	The inset in panel (a) shows the geometry assumed as a basic state for the linear stability analysis and for the derivation of the theoretical hillslope profiles (see also Methods).}
	\label{fig:stability}
\end{figure}

\begin{figure}
	\centering
	\includegraphics[trim={0.5cm 0.5cm 0.5cm 0cm},width=\textwidth]{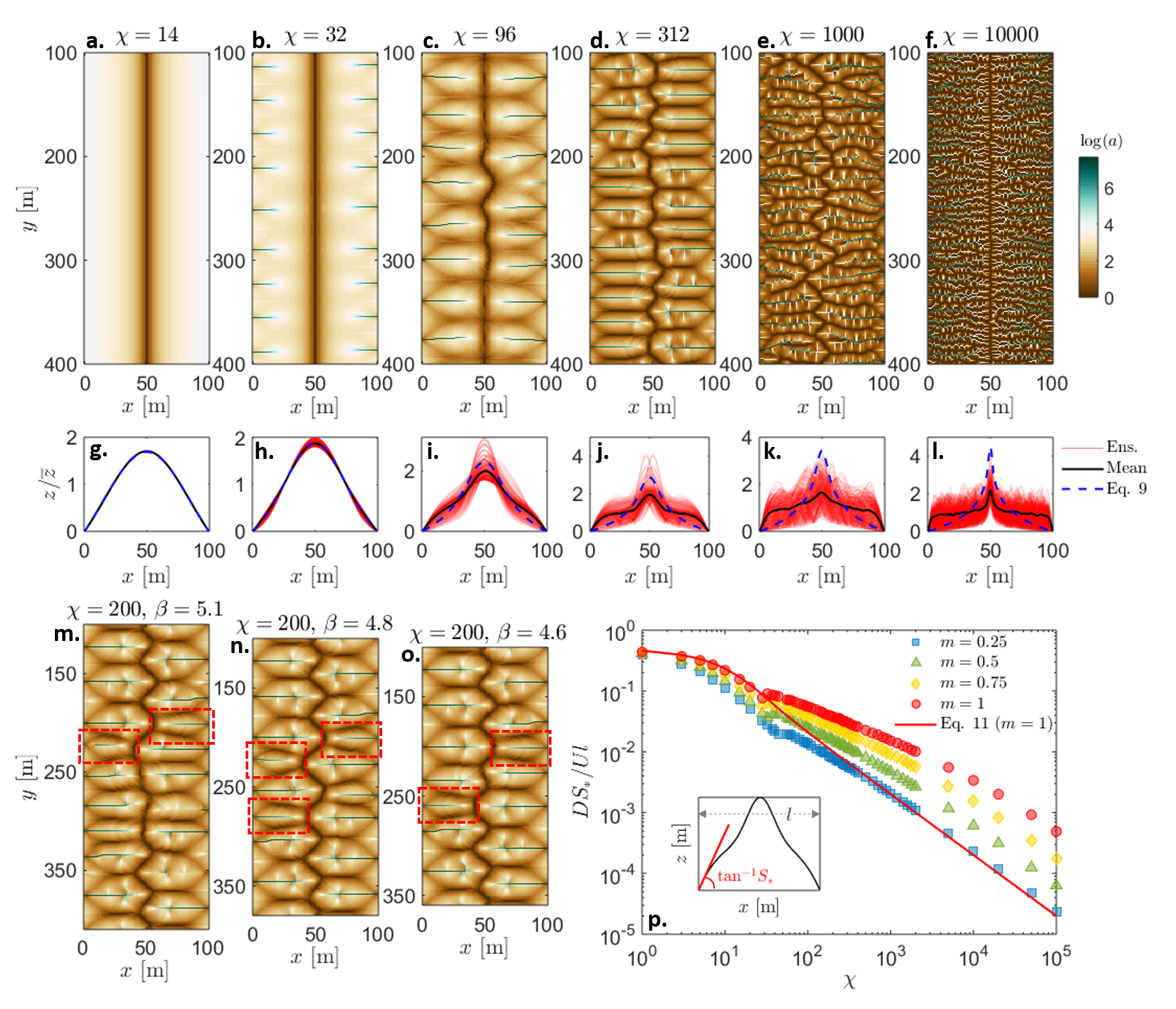}
	\caption{\textbf{Rectangular domains.} Ridge/valley networks obtained for $m=n=1$ over rectangular domains  with (a-f)  $\beta=5$ ($\chi$ = 14, 32, 96, 312, 1000, and 10000), (m) $\beta=5.1$ ($\chi$ = 200), (n) $\beta=4.8$ ($\chi$ = 200), and (o) $\beta=4.6$ ($\chi$ = 200). $\beta$ is a shape factor defined as the ratio between the two horizontal length scales $l_y$ and $l_x$, namely $\beta=l_y/l_x$. Examples of dislocation defects are shown by the red dashed rectangles in panels m-o. (g-l) Normalized elevation profiles along the $x$ axis for the six simulations of panels a-f: black lines are the mean elevation profiles, red lines show the ensemble of all the profiles along $x$, blue dashed lines are analytical elevation profiles for the unchannelized case -- equation \eqref{eq:z0}. Mean elevation profiles along the $x$ axis were computed as average values of the elevation profiles neglecting the extremal parts (100 m length) of the domain. (p) Slope of the mean elevation profile $S_*$  as a function of $\chi$ for simulations with $n=1$ and $m$ = 0.25, 0.5, 0.75, and 1. The solid red line represents the analytical solution for $m=1$ (equation \eqref{eq:S*}) for the unchannelized case. The schematic in the inset shows the definition of $S_*$ and $l$ used in the vertical axis of the chart.}
	\label{fig:rect1}
\end{figure}
\paragraph{Wide rectangular domains.}
To assess boundary-condition effects on branching patterns we also considered very wide rectangular domains ($\chi$ is constructed using the distance between the longest sides). 
Besides numerical investigation, in this case an analytical solution is possible for the unchannelized case (for $m=1$ and $n=1$, see Methods), on which bases we also performed a linear stability analysis. In our analogy with turbulent flows, this case corresponds to the flow of viscous fluids between parallel plates \citep{Drazin2004,Kundu2011}.

Results from the linear stability analysis are shown in Fig. \ref{fig:stability}. A critical value $\chi_{c} \approx 37$ for the first channel instability is identified, corresponding to a characteristic valley spacing $\lambda_c$ of approximately  42 m.  As $\chi$ further increases (i.e., runoff erosion increases with respect to diffusion) the predicted valley spacing is reduced (see Fig. \ref{fig:stability}c), with the formation of progressively narrower valleys. Results from the linear stability analysis are in line with predictions from numerical experiments conducted over large rectangular domains, where the first channel instability occurs at $\chi_n \approx 32$ with a valley spacing $\lambda_n \approx 33$ m. Analogously to the Orr-Sommerfeld problem for plane Poiseuille flow, the system here presents a Type I linear instability\cite{Cross1993}.

The numerical simulations confirm the results of the linear stability analysis and are in agreement with those of ref. \citenum{Perron2008}. Fig. \ref{fig:rect1} compares the drainage patterns obtained as a function of $\chi$ for rectangular domains of size 100 m by 500 m. As for the square domain, for small $\chi$ values the soil creep component dominates resulting in an unchannelized smooth topography (Fig. \ref{fig:rect1}a). After the first channelization, valleys tend to narrow as $\chi$ increases until the first secondary branching occurs (Fig. \ref{fig:rect1}b,c). Further increasing the runoff erosion component provides progressively more dissected landscapes with the emergence of secondary branching (Fig. \ref{fig:rect1}d-f). As in turbulent flows larger Reynolds numbers produce smaller and smaller vortices, here increasing $\chi$ leads to finer and finer branching (the resolution of which becomes quickly prohibitive from a computational standpoint).

The mean elevation profiles, computed as average elevation values along the $x$ axis and neglecting the terminal parts of the domain to avoid boundary effects, are shown in Fig. \ref{fig:rect1}g-l. As the topography becomes progressively more dissected with increasing $\chi$, the mean elevation profile tends to become more uniform (Fig. \ref{fig:rect1}g-l). Such a behavior of the mean elevation profiles for increasing $\chi$ is similar to the flattening of turbulent mean velocity profiles with increasing Reynolds number \citep{Kundu2011}.

The transition from a smooth to a channelized topography  with increasing $\chi$ is reflected in the behavior of the quantity $DS_*/Ul=f(\chi,m)$, which describes the ratio of the outgoing diffusive flux and the incoming uplift sediment flux at the hillslope base, $S_*$ being the slope of the mean elevation profile at the hillslope base (see Methods for details). Fig. \ref{fig:rect1}p shows the relationship between $DS_*/Ul$ and $\chi$ obtained from numerical simulations for $n=1$ and different values of the exponent $m$. For small $\chi$ values the numerical results match the analytic relationship for the smooth surface (equation \eqref{eq:S*}) and deviate from it at $\chi_n \approx 32$ where the first channel-forming instability occurs. Continuing our analogy with turbulence, the behavior of $DS_*/Ul$ as a function of $\chi$ closely resembles that of the friction factor with increasing Reynolds number (see Methods as well as Figure 7.3 in ref. \citenum{Panton1984}).

The effect of boundary conditions on the spatial regularity of ridge and valley patterns becomes especially apparent when comparing simulations with different aspect ratios. As can be seen in Fig.  \ref{fig:rect1}m-o, when the domain size is slightly changed, the spatial organization of ridges and valleys is modified (see, e.g., the more regular pattern obtained for $\beta=4.6$ compared to $\beta=5.1$), while the mean elevation profiles remain practically invariant (Fig. \ref{figS8:defects}). This suggests that some optimal domain length is needed to accommodate the formation of regular ridge and valley patterns (this is also evident from an analysis of cross-sections along the longer sides of the domain in Figs. \ref{figSI:Sect1}-\ref{figSI:Sect5}). This results in the formation of dislocation defects, as highlighted in the example of Fig. \ref{fig:rect1}m-o, as it is typical in nonlinear pattern-forming PDEs \citep{Cross1993}.

\section*{Discussion and conclusions}
A succession of increasingly complex networks of ridges and valleys was produced by a system of nonlinear PDEs serving as a minimalist model for landscape evolution in detachment-limited conditions. The sequence of instabilities is reminiscent of the subsequent bifurcations in fluid dynamic instabilities \citep{Cross1993,Drazin2004,Kundu2011} and is captured by a dimensionless number ($\chi$) accounting for the relative importance of runoff erosion, soil creep, and uplift in relation to the typical domain size. Tantalizing analogies with fluid turbulence, and in general with other driven  non-equilibrium systems in which a hierarchical pattern develops toward finer scales, can also be observed in the competition between runoff erosion and soil creep (which resembles the competition between viscous and inertial forces), the reduction of the minimum branching scale with $\chi$, and the flattening of the mean hypsometric curves as the channelization is increased.

Characteristic spatial configurations were shown to emerge over both square and rectangular domains from the trade-off between diffusion and erosion. The striking regularity of the ridge and valley networks, with the characteristics of regular pre-fractals (e.g., the Peano basin \citep{Mandelbrot1982,Marani1991,Rodriguez1992,Flammini1996}), is quickly lost as effects of noise and irregular boundaries are introduced. The shape of the hypsometric curve depends on the level of channelization and branching \citep{Willgoose1998} and thus on the dominant erosional mechanisms acting on the landscape (i.e., interplay between runoff erosion, soil creep, and uplift) and the various landscape properties affecting diffusion and erosion coefficients, such as soil type, vegetation cover, and climate. When diffusion dominates,  hypsometric curves display a less pronounced basal convexity \citep{Willgoose1998}. 

Future work will focus on transient dynamics to explore the differences between the hypsometry of juvenile and old landscapes. It is likely that, during the early stages of the basin development when the drainage network is formed, the hypsometric curve will present a more pronounced basal convexity \citep{Strahler1952} regardless of the value of $\chi$, progressively transitioning toward its quasi-equilibrium form during the ``relaxation phase'' \citep{Bonetti2017GRL}. It will be interesting to compare such slow relaxations (e.g., Fig. \ref{fig:rect1}), often towards slightly irregular configurations rather than perfectly regular networks, with the presence of defects in crystals and the amorphous configurations originating in glass transition \citep{Debenedetti2001}.

\section*{Methods}
\textbf{Analytical solutions for $m=n=1$.} To derive one-dimensional steady state solutions of the coupled PDE system (equations \eqref{eq5:a}-\eqref{eq5:z3}) we consider a symmetric hillslope of length $l$ in the $x$-direction, with divide at $x=0$ (see inset in Fig. \ref{fig:stability}a). Assuming a fixed elevation $z=0$ at $x=\pm l/2$, the steady steady solution of the coupled system \eqref{eq5:a}-\eqref{eq5:z3} for $m=n=1$ reads
\begin{eqnarray}
    a_0&=& |x| \label{eq:a0}\\
    z_0&=& \frac{U}{2D} \left[\left(\frac{l}{2}\right)^2 \mathcal{H}\left(1,1;\frac{3}{2},2;-\frac{K_a \left(\frac{l}{2}\right)^2}{D}\right)-x^2 \mathcal{H}\left(1,1,;\frac{3}{2},2;-\frac{K_a x^2}{D} \right) \right] \label{eq:z0}
\end{eqnarray}
where subscript 0 denotes the basic steady state, and $\mathcal{H}(.,.;.,.;.)$ is the generalized hypergeometric function \cite{Abramowitz1964}. In these conditions, the local slope $S_0=dz_0/dx$ can also be derived analytically as
\begin{eqnarray}
    S_0&=& \frac{\sqrt{2}U \mathcal{D}\left(\frac{\sqrt{K_a}x}{\sqrt{2D}}\right)}{\sqrt{D K_a}} \label{eq:S0}
\end{eqnarray} 
where $\mathcal{D}(.)$ is the Dawson's integral \cite{Abramowitz1964}.
\\
\\
\textbf{Linear stability analysis.}
We studied the stability of the basic state (equations \eqref{eq:a0}-\eqref{eq:z0}) to perturbations $\Tilde{a}$ and $\Tilde{z}$ in the $y$-direction. Boundary conditions are zero sediment and specific drainage area at the hilltop ($\Tilde{a}=d\Tilde{z}/dx=0$ at $x=0$) and fixed elevation at the domain boundary ($\Tilde{z}=0$ at $x=l/2$). We use normal mode analysis and write perturbations in the classical form $\Tilde{a}=\phi(x) e^{iky+\sigma t}$ and $\Tilde{z}=\psi(x) e^{iky+\sigma t}$  (plus complex conjugate), where $k$ and $\sigma$ are the wavenumber and the growth rate of the perturbations, respectively. The perturbed system can be re-cast in terms of a third order non-constant coefficient differential eigenvalue problem of the form $\gamma_1(x)\phi'''(x)+\gamma_2(x)\phi''(x)+\gamma_3(x)\phi'(x)+\gamma_4(x)\phi(x)=\sigma \gamma_5(x) \phi'(x)$. Solutions to the stability problem are obtained by means of a spectral Galerkin technique with numerical quadrature \citep{Canuto2006,Camporeale2012}. Among the discrete set of eigenvalues obtained, we tracked the behavior of
the least stable (i.e., with largest real part).
\\
\\
\textbf{Numerical simulations.} 
Numerical simulations were performed using forward differences in time and centered difference approximations for the spatial derivatives, considering regular square grids of lateral dimension $l$, as well as on rectangular domains with shape factor $\beta$, defined as the ratio between the domain dimensions in the $y$ and $x$ direction (i.e., $\beta=l_y/l_x$). Specifically, in the simulations over rectangular domains we fixed the length in the $x$ direction (i.e., $l_x=100$ m), and varied only the length $l_y$ in the $y$ direction.
The total drainage area $A$ was computed at each grid point with the $D\infty$ algorithm, while $a$  was then approximated as $A/\Delta x$  (ref. \citenum{Tarboton1997}), with $\Delta x$ the grid size.Simulations were run assuming $\Delta x=1$ m (additional numerical experiments, shown in Fig. \ref{figS1:gridsize},  were performed for different grid sizes to validate the independence of the resulting patterns on the grid resolution).
Convex profiles were used as initial condition. Over wide rectangular domains for $\chi \ge 320$ a white noise with standard deviation equal to 10$^{-6}$ m was also added in the initial condition. A sensitivity analysis was conducted over square domains (not shown) to make sure that the resulting spatial organization of ridges and valleys at steady state was robust to the choice of initial conditions.
We considered a wide range of $\chi$ values (from 10$^0$ to 10$^5$) constructed by using literature values of the system parameters, which are generally estimated in terms of time-averaged values from experimental hillslope shapes \citep{Sweeney2015} or high resolution topographies \citep{Perron2008,Perron2012}.

\textbf{Dimensional analysis of the channelization transition.}
In channel and pipe flows the relationship between the friction factor $\xi$ and the Reynolds number $Re$ can be obtained by first relating the wall shear stress $\tau=\mu  d\overline{u}/dx^*|_{x^*=0}$, where $\overline{u}$ is the streamwise mean velocity profile and $x^*$ is the distance from the wall, to 
its governing quantities as  $\tau=\Xi(V,l,\mu,\rho,\epsilon)$, where $\rho$ is the density, $\mu$ the viscosity, $V$ the mean velocity, $l$ the characteristic lateral dimension, and $\epsilon$ the roughness height. The Pi-Theorem then may be used to express the head loss per unit length ($g$ is gravitational acceleration) as $S_h=\frac{4\tau}{g \rho l}=\frac{V^2}{2gl}\xi\left(Re,\frac{\epsilon}{l}\right)$, see Ref. \citenum{Munson1995}. Analogously, here we can relate the slope of the mean elevation profile at the hillslope base $S_*=d\overline{z}/dx|_{x=l/2}$ to the parameters and characteristics of the landscape evolution model as $S_*=\Phi(D,K_a,m,U,l)$ (we consider here $n=1$). Choosing $l$, $U$, and $D$ as dimensionally independent variables, the Pi-Theorem yields $DS_*/Ul=\varphi(\chi, m)$, where the quantity $DS_*$ quantifies the diffusive outgoing sediment flux per unit width (along the $x$-axis) at the boundary, while the term $Ul$ represents the incoming sediment flux by tectonic uplift per unit width. Such a functional relationship can be analytically derived for the unchannelized case when $m=1$ from (\ref{eq:S0}) as
\begin{equation}
    \frac{DS_*}{Ul}=\left(\frac{\chi}{2}\right)^{-1/2}\mathcal{D}\left[ \left(\frac{\chi}{8}\right)^{1/2}\right].
    \label{eq:S*}
\end{equation}
In the numerical simulations, $S_*$ was computed as the slope of the linear fit to the mean elevation profile in the first 3 meters at the hillslope base (see inset in Fig. \ref{fig:rect1}p).
\\
\paragraph{Acknowledgements} We acknowledge support from the US National Science Foundation (NSF) grants EAR-1331846 and EAR-1338694, and BP through the Carbon Mitigation Initiative (CMI) at Princeton University.
\paragraph{Author Contribution} S.B. and A.P. designed research, discussed results, and wrote the paper. S.B. and M.H. performed the numerical simulations, while S.B., C.C., and A.P. performed the linear stability analysis. All the authors reviewed and edited the final version of the manuscript.
\paragraph{Competing interests} The authors declare no competing interests.
\bibliographystyle{unsrt}
\bibliography{Biblio}

\newpage
\section*{\centering{SUPPLEMENTARY INFORMATION}}

\textbf{Introduction}\\
Additional results from the landscape evolution model are presented in this Supplementary Information. In particular, a sensitivity analysis was performed to investigate the effect of grid size on the drainage patterns obtained from the numerical experiments (results are shown in Fig. \ref{figS1:gridsize} and discussed below).
Hypsometric curves obtained for different values of the water flow exponent $m$ and slope exponent $n$ are shown in Fig. \ref{figS3:changeexponentshypso}. Additional results for rectangular domains are provided in Figs. \ref{figSI:Sect1}-\ref{figS8:defects} and discussed below.
\\
\textbf{Effect of grid resolution} \\
As explained in the main text, using the specific drainage area $a$ as a proxy for the water flux ensures the modeled ridge/valley patterns to be robust to grid size variations, while theoretically justified in terms of a continuity equation.
To assess the independence of such drainage patterns on the specific grid resolution chosen in the simulations, additional numerical tests were performed. In particular, simulations were run over the square domain of size $l=100$ m for $\chi$ = 40, 125, and 200 assuming grid sizes $\Delta x$ equal to 0.5 and 2 m. Results are compared to those obtained for $\Delta x=1$ m in Fig. \ref{figS1:gridsize}, confirming that the resulting patterns and hypsometric curves are independent of the grid size value $\Delta x$.
\\
\textbf{Results for rectangular domains}\\
Figs. \ref{figSI:Sect1}-\ref{figSI:Sect5} show cross-sections along the longer side of  rectangular domains for different values of the dimensionless parameter $\chi$. It is interesting to note that, while the probability distributions of the elevation profiles are similar between cross-sections on each side of the main central ridgeline (blue and black, respectively), the lack of spatial synchronicity of ridges and valleys in the cross sections suggests that a proportionality between the domain dimension and the characteristic valley spacing is needed to accomodate the formation of regular ridge/valley patterns.

Fig. \ref{figS8:defects} compares the mean elevation profiles obtained for $\chi=200$ and shape factor $\beta$ equal to 4.6, 4.8, and 5.1 (i.e., $l_x= 100$ m and $l_y=$ 460, 480, and 510 m in the three cases) -- ridge and valley patterns for these simulations are shown in Fig. 4m-o in the main text.

\setcounter{figure}{0}
\makeatletter 
\renewcommand{\thefigure}{S\@arabic\c@figure}
\makeatother

\begin{figure}[h!]
\centering
\centerline{\includegraphics[trim={0.5cm 0.5cm 0.5cm 0.5cm},width=.8\columnwidth]{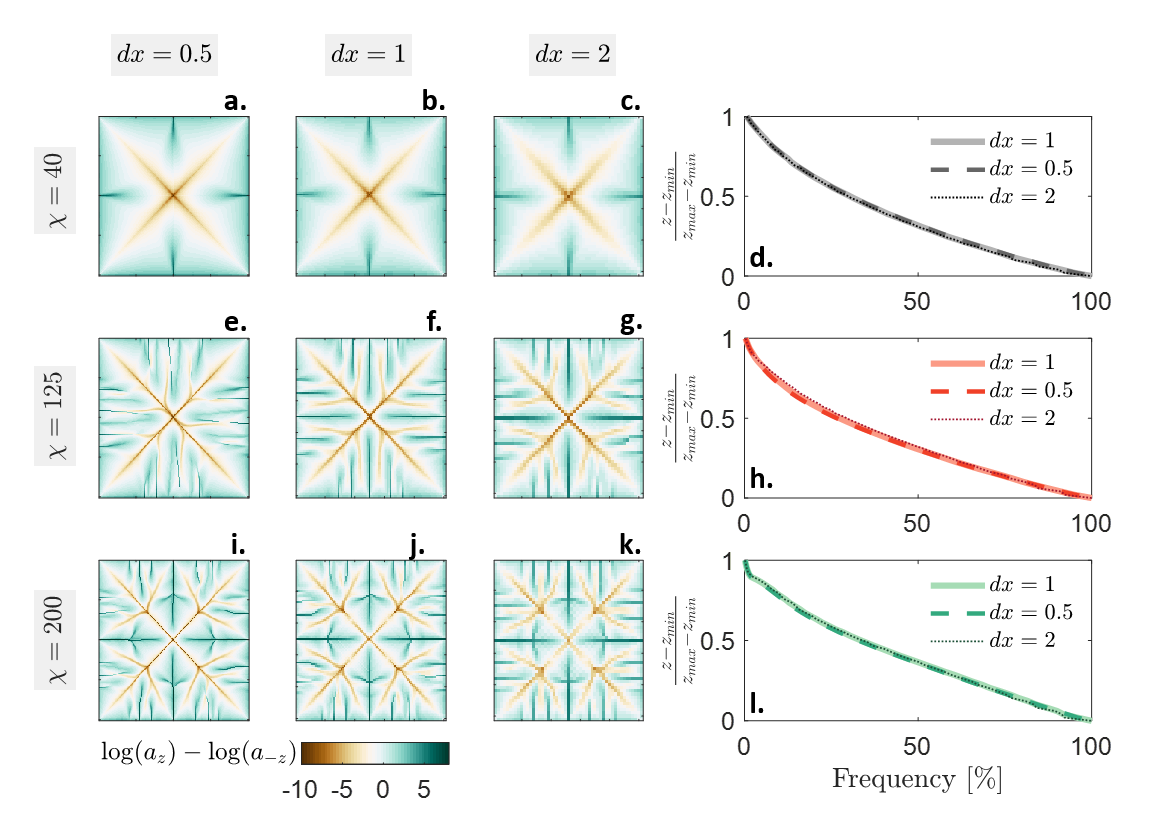}}
\caption{Ridge/valley patterns (brown = ridge, green = valley) for $\chi$ = 40 (a-c), 125 (e-g), and 200 (i-k) and $\Delta x$ = 0.5 (a, e, i), 1 (b, f, j), and 2 (c, g, k). The corresponding normalized hypsometric curves are shown in panel d, h, and l for $\chi$ = 40, 125, and 200, respectively.}
\label{figS1:gridsize}
\end{figure}

\begin{figure}
\centering
\centerline{\includegraphics[trim={0cm 0cm 0cm 0cm},width=\columnwidth]{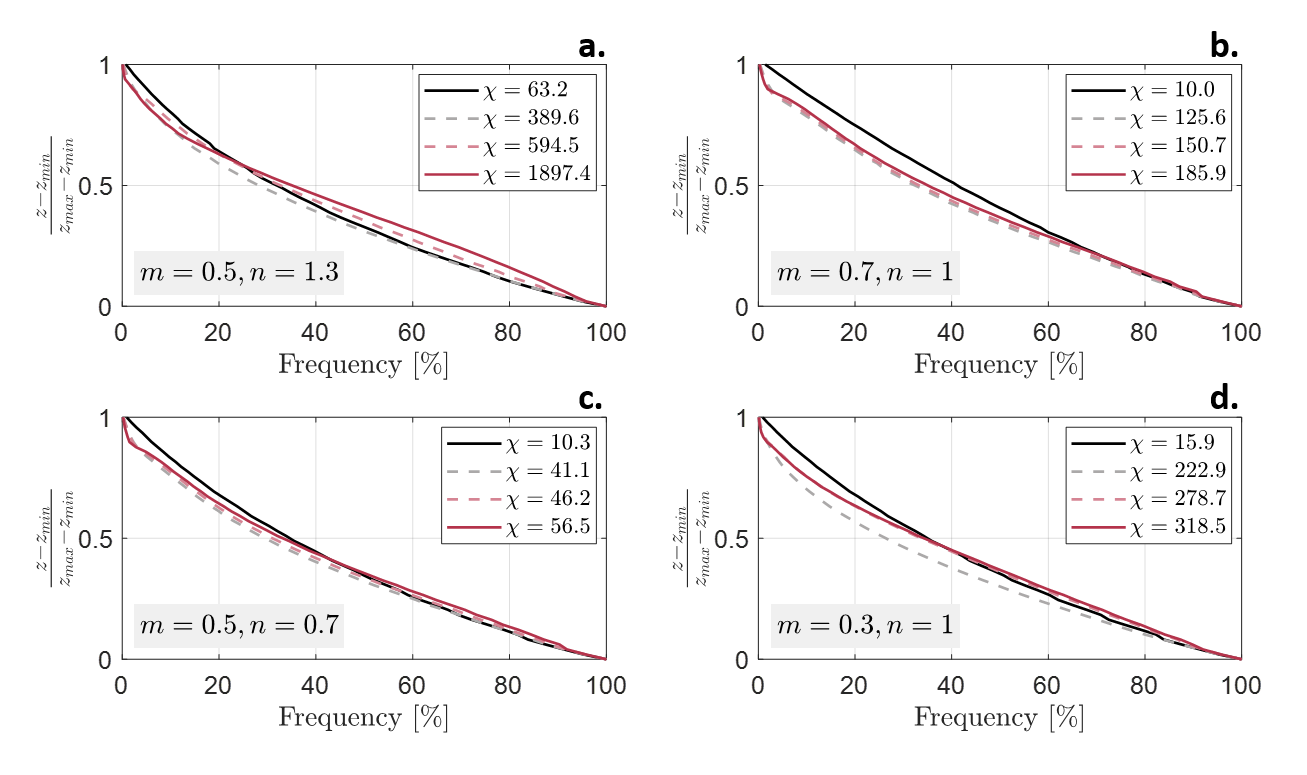}}
\caption{Normalized hypsometric curves obtained for different values of the slope and runoff exponents (i.e., $n$ and $m$). The black line represents the smooth solution with no channels, the dashed gray lines represents cases in which channels are formed but no secondary branching is observed, while dashed and solid red lines show results with secondary branching. 2d representations of the planar geometric patterns of ridges and valleys for these cases are shown in the main text (Fig. 2).}
\label{figS3:changeexponentshypso}
\end{figure}

\begin{figure}
\centering
\includegraphics[trim={0cm 0cm 0cm 0cm},width=.9\columnwidth]{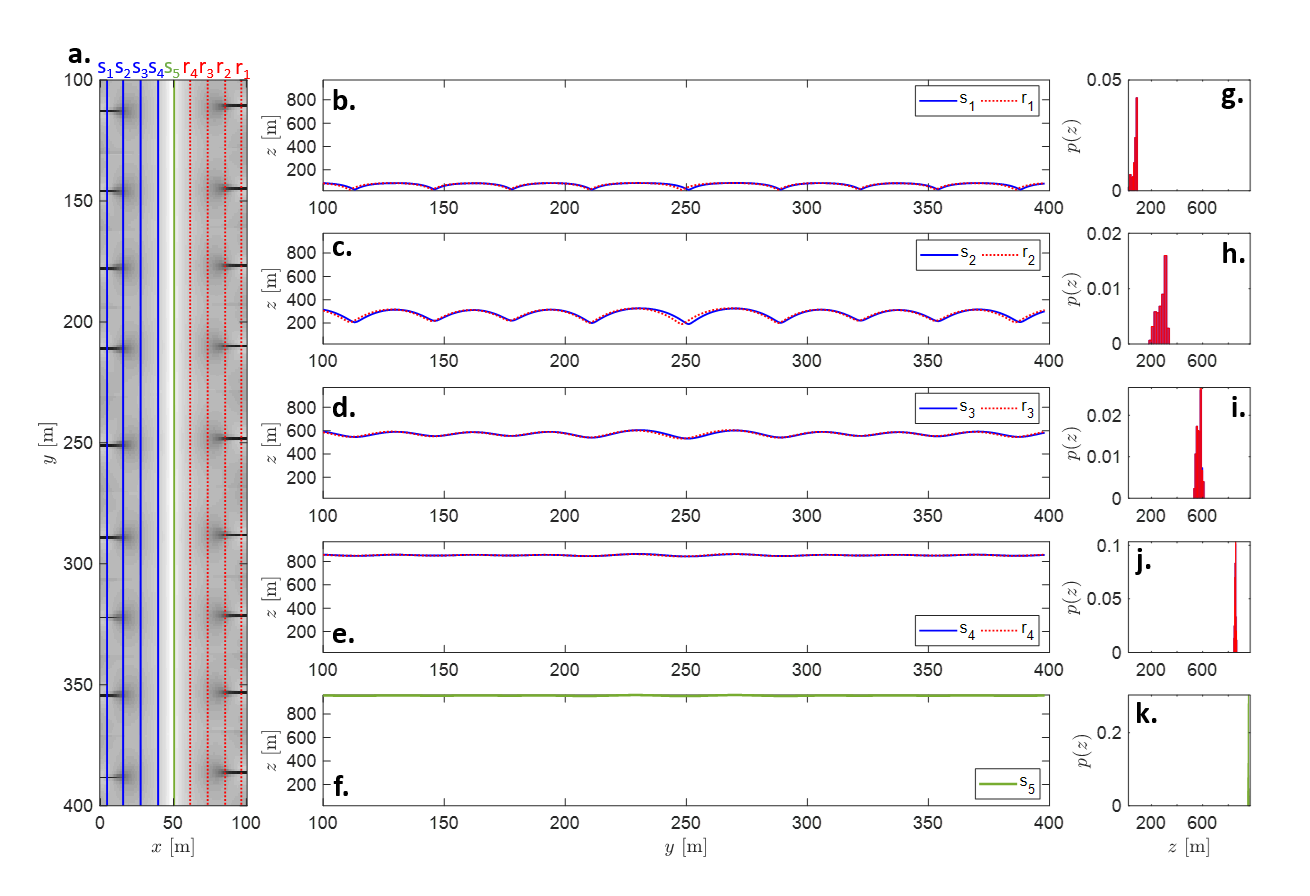}
\caption{(b-f) Cross-sections of the topographic surface obtained for $\chi=32$ and $m=n=1$ over a rectangular domain of size 500m x 100m ($\beta=5$, panel a). Probability distributions of the elevation profiles along the cross-sections are shown in panels g-k. Blue and red colors refer to cross sections on the left and right side of the domain, respectively, while green refers to the cross section along the central ridgeline at $x = 50$ m (see location of the transects in panels a).}
\label{figSI:Sect1}
\end{figure}

\begin{figure}
\centering
\includegraphics[trim={0cm 0cm 0cm 0cm},width=.9\columnwidth]{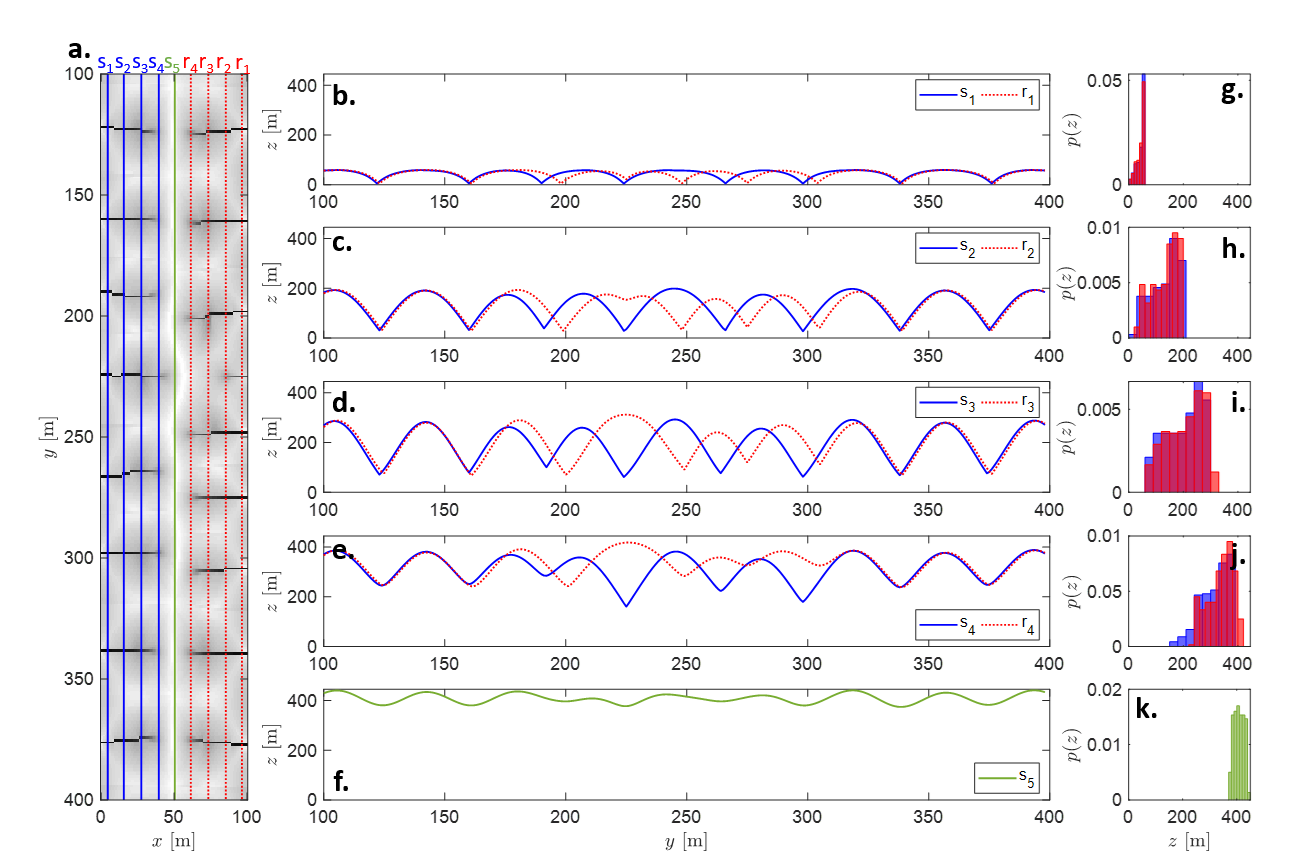}
\caption{(b-f) Cross-sections of the topographic surface obtained for $\chi=96$ and $m=n=1$ over a rectangular domain of size 500m x 100m ($\beta=5$, panel a). Probability distributions of the elevation profiles along the cross-sections are shown in panels g-k. Blue and red colors refer to cross sections on the left and right side of the domain, respectively, while green refers to the cross section along the central ridgeline at $x = 50$ m (see location of the transects in panel a).}
\label{figSI:Sect2}
\end{figure}

\begin{figure}
\centering
\includegraphics[trim={0cm 0cm 0cm 0cm},width=.9\columnwidth]{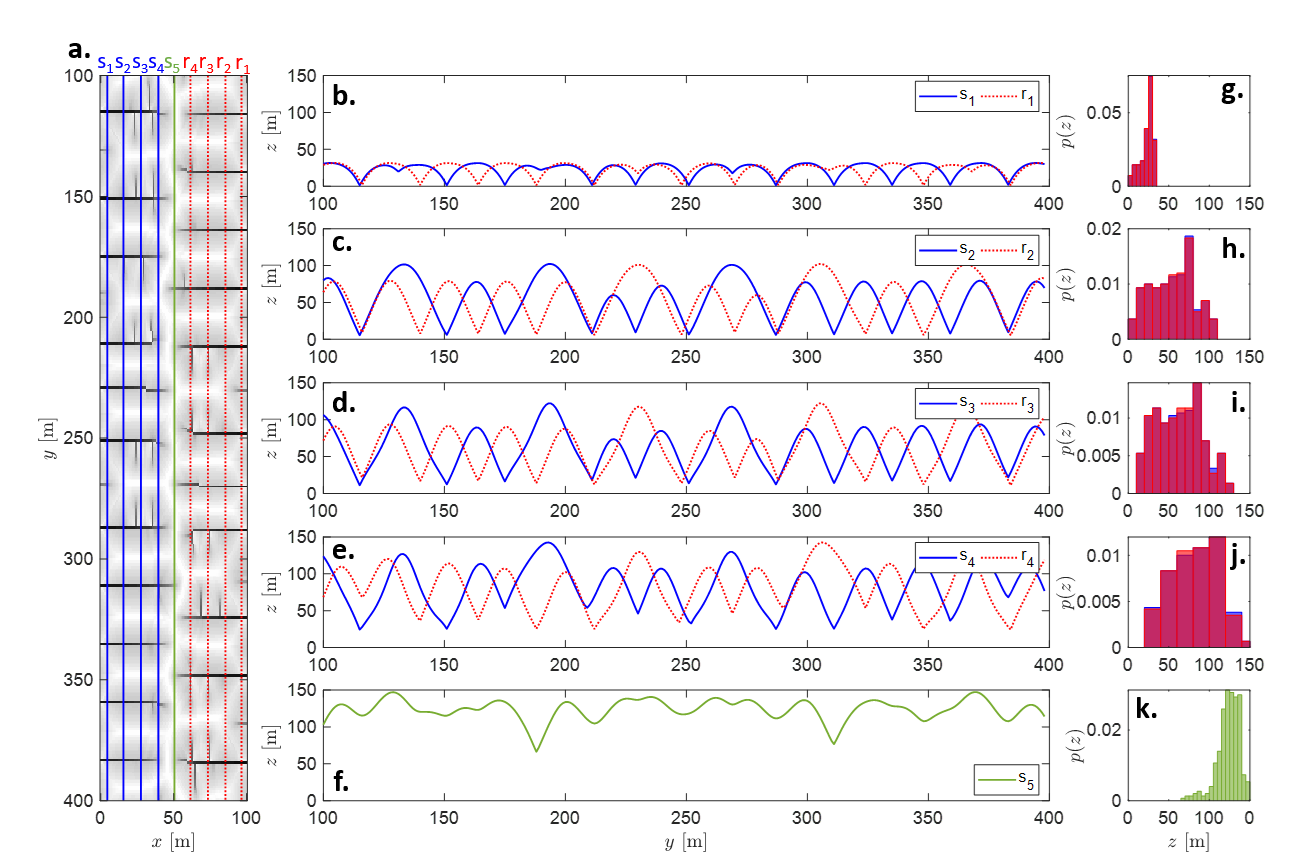}
\caption{(b-f) Cross-sections of the topographic surface obtained for $\chi=312$ and $m=n=1$ over a rectangular domain of size 500m x 100m ($\beta=5$, panel a). Probability distributions of the elevation profiles along the cross-sections are shown in panels g-k. Blue and red colors refer to cross sections on the left and right side of the domain, respectively, while green refers to the cross section along the central ridgeline at $x = 50$ m (see location of the transects in panel a).}
\label{figSI:Sect3}
\end{figure}

\begin{figure}
\centering
\includegraphics[trim={0cm 0cm 0cm 0cm},width=.9\columnwidth]{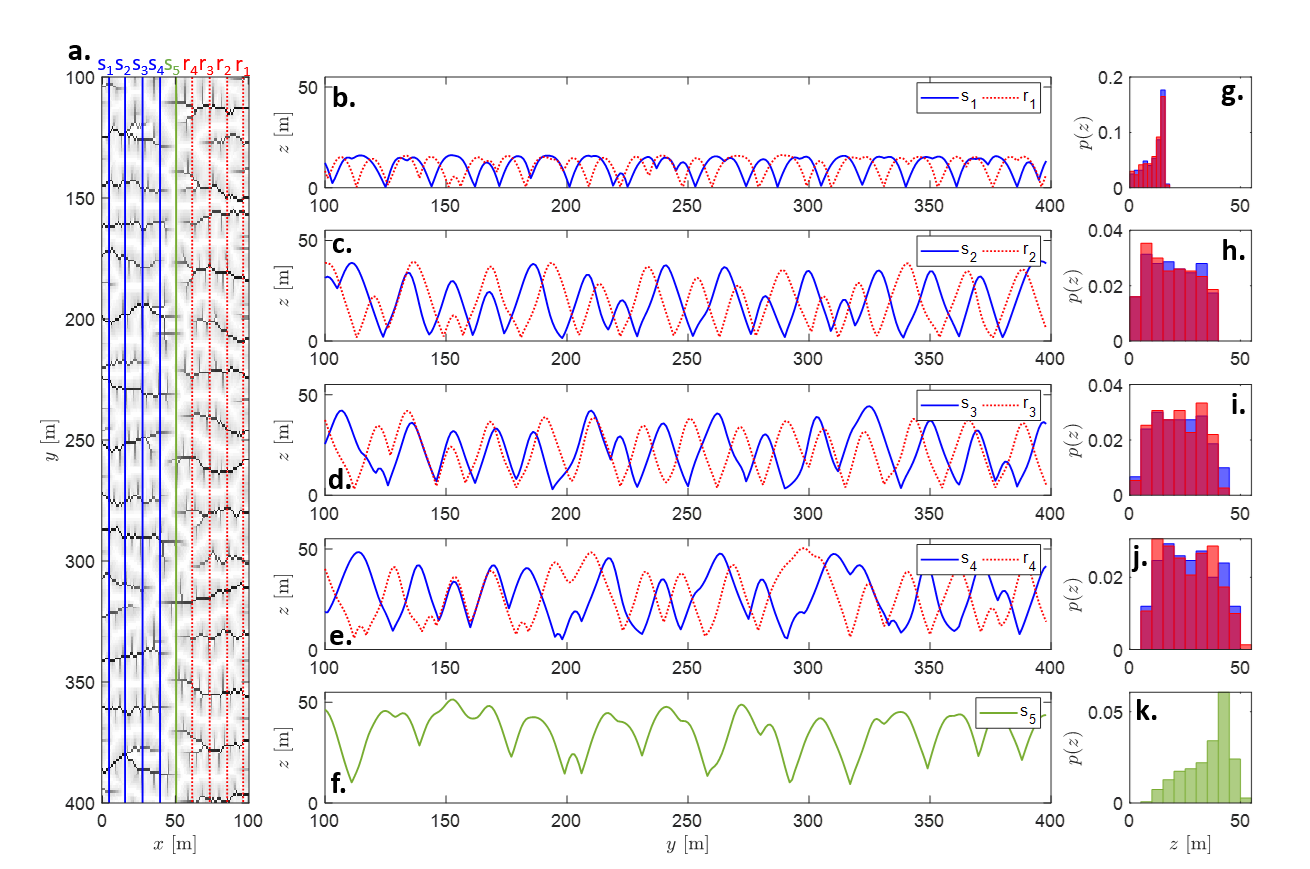}
\caption{(b-f) Cross-sections of the topographic surface obtained for $\chi=1000$ and $m=n=1$  over a rectangular domain of size 500m x 100m ($\beta=5$, panel a). Probability distributions of the elevation profiles along the cross-sections are shown in panels g-k. Blue and red colors refer to cross sections on the left and right side of the domain, respectively, while green refers to the cross section along the central ridgeline at $x = 50$ m (see location of the transects in panel a).}
\label{figSI:Sect4}
\end{figure}

\begin{figure}
\centering
\includegraphics[trim={0cm 0cm 0cm 0cm},width=.9\columnwidth]{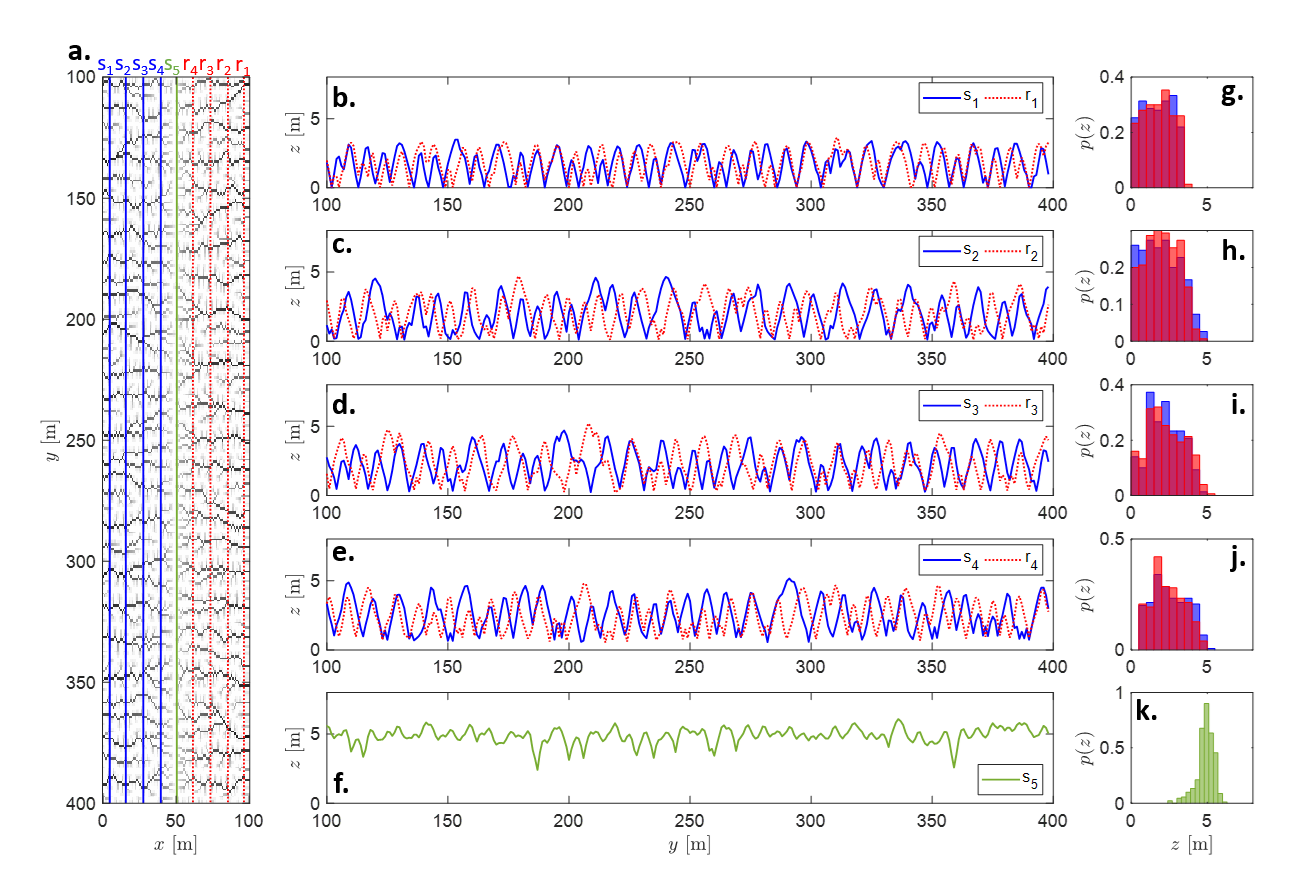}
\caption{(b-f) Cross-sections of the topographic surface obtained for $\chi=10000$ and $m=n=1$ over a rectangular domain of size 500m x 100m ($\beta=5$, panel a). Probability distributions of the elevation profiles along the cross-sections are shown in panels g-k. Blue and red colors refer to cross sections on the left and right side of the domain, respectively, while green refers to the cross section along the central ridgeline at $x = 50$ m (see location of the transects in panel a).}
\label{figSI:Sect5}
\end{figure}

\begin{figure}
\centering
\centerline{\includegraphics[trim={0cm 0cm 0cm 0cm},width=\columnwidth]{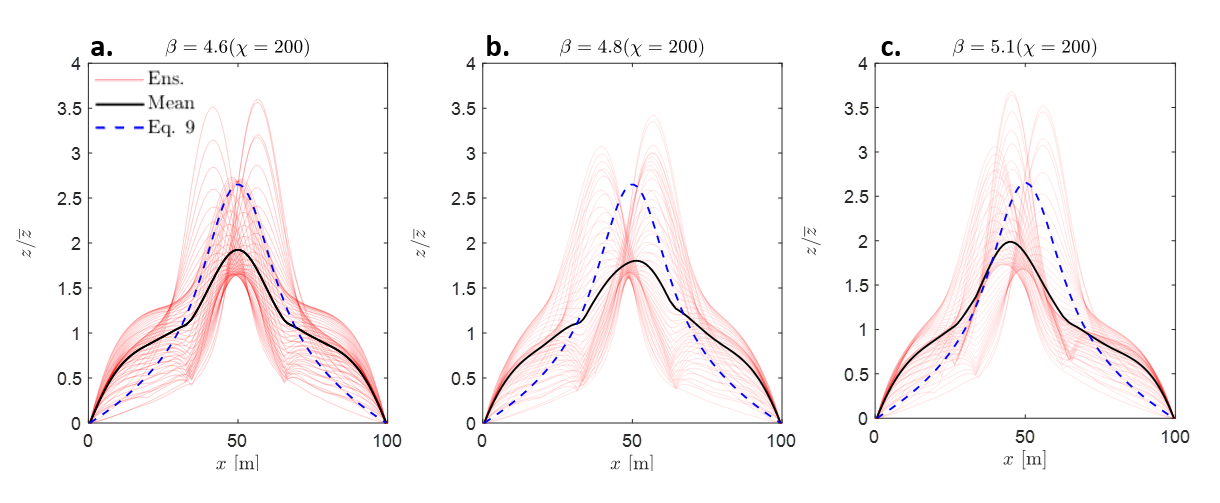}}
\caption{Normalized elevation profiles along the $x$ axis for the three simulations of Fig. 4m-o in the main text: (a) $\beta=4.6$, (b) $\beta=4.8$, (c) $\beta=5.1$ (all simulations have $m=n=1$ and $\chi=200$). Black lines are the mean elevation profiles, red lines show the ensemble of all the profiles along $x$, blue dashed lines are analytical elevation profiles for the unchannelized case (equation (9) in the manuscript). Mean elevation profiles along the $x$ axis were computed as average values of the elevation profiles neglecting the extremal parts (100 m length) of the domain.}
\label{figS8:defects}
\end{figure}

\end{document}